\documentclass{emulateapj}
\usepackage{graphicx}
\usepackage{epstopdf}
\def\ie{{\it i.e., }}
\def\be{\begin{equation}}
\def\ee{\end{equation}}
\def\bea{\begin{eqnarray}}
\def\eea{\end{eqnarray}}
\newcommand{\eg}{{\it e.g., }} 

\begin{document}

\title{Simultaneous Estimation of Photometric Redshifts and SED Parameters:\\ Improved Techniques and a Realistic Error Budget}

\author{Viviana Acquaviva\altaffilmark{1}, Anand~Raichoor\altaffilmark{2,3,4}, Eric Gawiser\altaffilmark{5}}

\shorttitle{Simultaneous Photo-z Estimation and SED fitting} 

\shortauthors{Acquaviva et al.} 

\altaffiltext{1}{Department of Physics, CUNY NYC College of Technology, 300 Jay Street, Brooklyn NY 11201; vacquaviva@citytech.cuny.edu}
\altaffiltext{2}{CEA, Centre de Saclay, Irfu/SPP, F-91191 Gif-sur-Yvette, France}
\altaffiltext{3}{GEPI, Observatoire de Paris, 77 av. Denfert Rochereau, 75014 Paris, France}
\altaffiltext{4}{INAF -- Osservatorio Astronomico di Brera, via Brera 28, I-20121 Milan, Italy}
\altaffiltext{5}{Department of Physics and Astronomy, Rutgers, The State University of New Jersey, 136 Frelinghuysen Road, Piscataway, NJ 08854}

\def\LaTeX{L\kern-.36em\raise.3ex\hbox{a}\kern-.15em
    T\kern-.1667em\lower.7ex\hbox{E}\kern-.125emX}

\label{firstpage}

\begin{abstract}

We seek to improve the accuracy of joint galaxy photometric redshift estimation and spectral energy distribution (SED) fitting. By simulating different sources of uncorrected systematic errors, we demonstrate that if the uncertainties on the photometric redshifts are estimated correctly, so are those on the other SED fitting parameters, such as stellar mass, stellar age, and dust reddening. Furthermore, we find that if the redshift uncertainties are over(under)-estimated, the uncertainties in SED parameters tend to be over(under)-estimated by similar amounts. These results hold even in the presence of severe systematics and provide, for the first time, a mechanism to validate the uncertainties on these parameters via comparison with spectroscopic redshifts. 
We propose a new technique (annealing) to re-calibrate the joint uncertainties in the photo-z and SED fitting parameters without compromising the performance of the SED fitting + photo-z estimation. This procedure provides a consistent estimation of the multidimensional probability distribution function in SED fitting + z parameter space, including all correlations. While the performance of joint SED fitting and photo-z estimation might be hindered by template incompleteness, we demonstrate that the latter is ``flagged" by a large fraction of outliers in redshift, and that significant improvements can be achieved by using flexible stellar populations synthesis models and more realistic star formation histories. In all cases, we find that the median stellar age is better recovered than the time elapsed from the onset of star formation.
Finally, we show that using a photometric redshift code such as EAZY to obtain redshift probability distributions that are then used as priors for SED fitting codes leads to only a modest bias in the SED fitting parameters and is thus a viable alternative to the simultaneous estimation of SED parameters and photometric redshifts.

\end{abstract}


\section{Introduction}

Spectral Energy Distribution fitting is a useful tool to infer the physical properties of galaxies, such as mass, stellar age, star formation history, and dust content, starting from photometric observations.  The basic idea is to use Stellar Population Synthesis (SPS) models to create templates whose properties are known, and then explore the parameter space of possible properties to find models that resemble the data. The success rate of this technique relies crucially on the accuracy of the chosen templates, while its efficiency depends mostly on the algorithm used for the exploration of the parameter space. 

A problem that is fundamentally related to SED fitting is the determination of the redshift of galaxies (called photo-z) on the basis of the same photometric data. This problem will become increasingly important since the large imaging surveys of the future (\eg the Large Synoptic Survey Telescope, \citealt{LSST}) will provide photometric broad-band coverage for a very large number of galaxies, but spectroscopic follow-up will only be available for a small fraction of them. 

There are a number of publicly available algorithms to estimate photo-z (see \eg the discussion in \citealt{Walcher2011} and http://www.sedfitting.org/SED08/Fitting.html for a complete list). These codes use linear combinations of empirical templates or SPS templates (e.g., EAZY, \citealt{EAZY}; HYPERZ, \citealt{HYPERZ}), or adopt supervised machine learning techniques (\eg AnnZ, \citealt{AnnZ}; TPZ, \citealt{TPZ}) to optimize the estimation of the photometric redshifts. 

A common practice is to perform the estimation of the photometric redshifts and the SED fitting in two steps, by determining the photo-z first, and then fixing the redshift at the best-fit value to estimate the other physical properties of the galaxy. However, this procedure causes an underestimation of the uncertainties in the SED fitting parameters, since the uncertainty in redshift is not propagated correctly. Even when the latter is taken into account in the SED fitting algorithm, parameter estimation might still be biased because the templates used in the photo-z estimation and SED fitting differ, and if the SED fitting code doesn't utilize the full information from the redshift probability distribution (see \eg \citealt{Kotulla2013,Pirzkal2013}). A different example of parameter estimation where the results of a two-step process differ significantly from the fully covariant exploration of parameter space is discussed in \cite{Andreon2012}. Here we seek to develop a strategy to estimate photometric redshift and perform SED fitting simultaneously, by using the same algorithm and the same templates, and to improve the accuracy of the error estimation on redshift and other physical properties of galaxies by using a novel technique ({\it annealing}) to calibrate the uncertainties.

\begin{figure*}[t]
\begin{centering}
\includegraphics[width=0.49\linewidth]{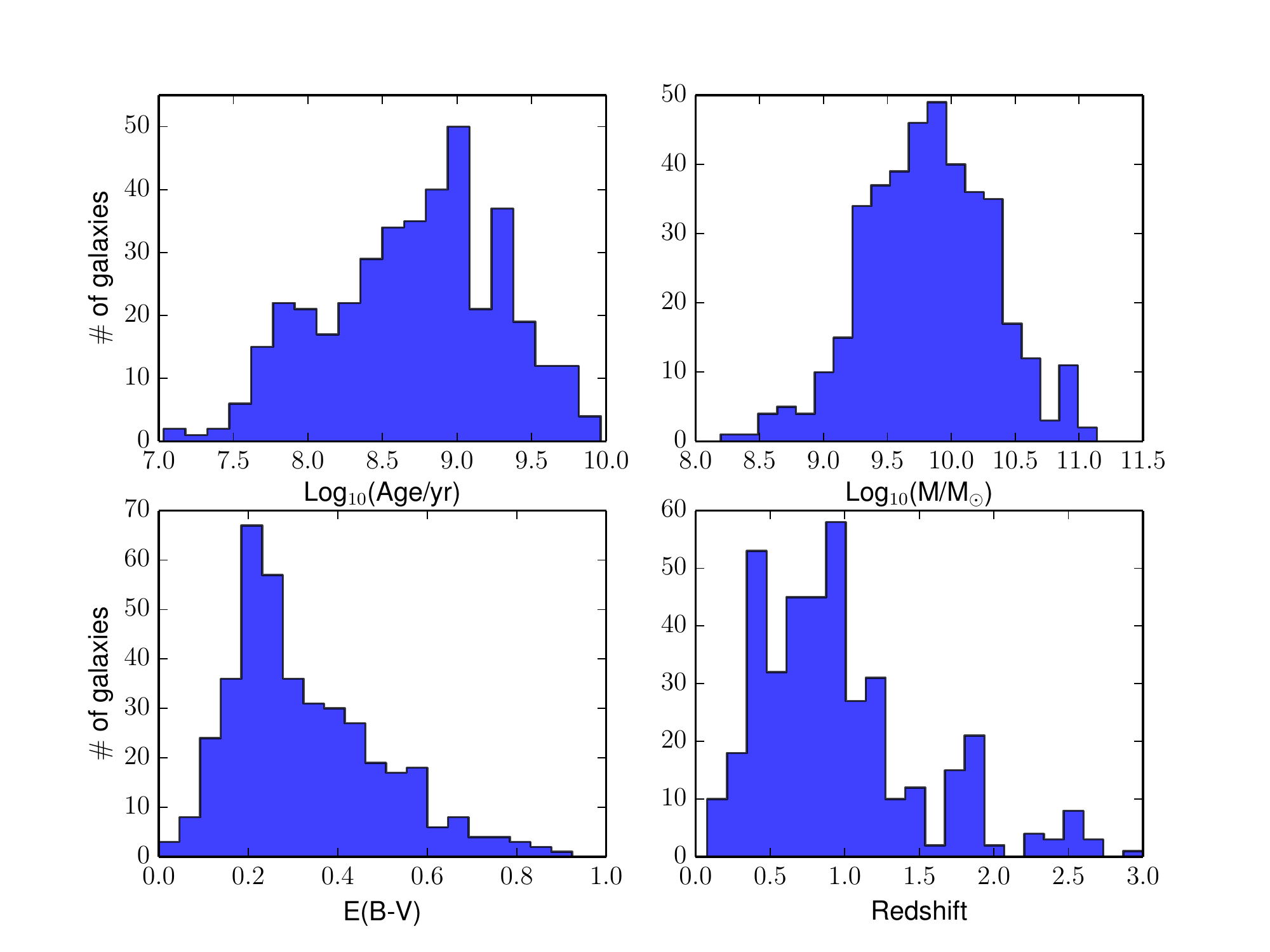}
\includegraphics[width=0.49\linewidth]{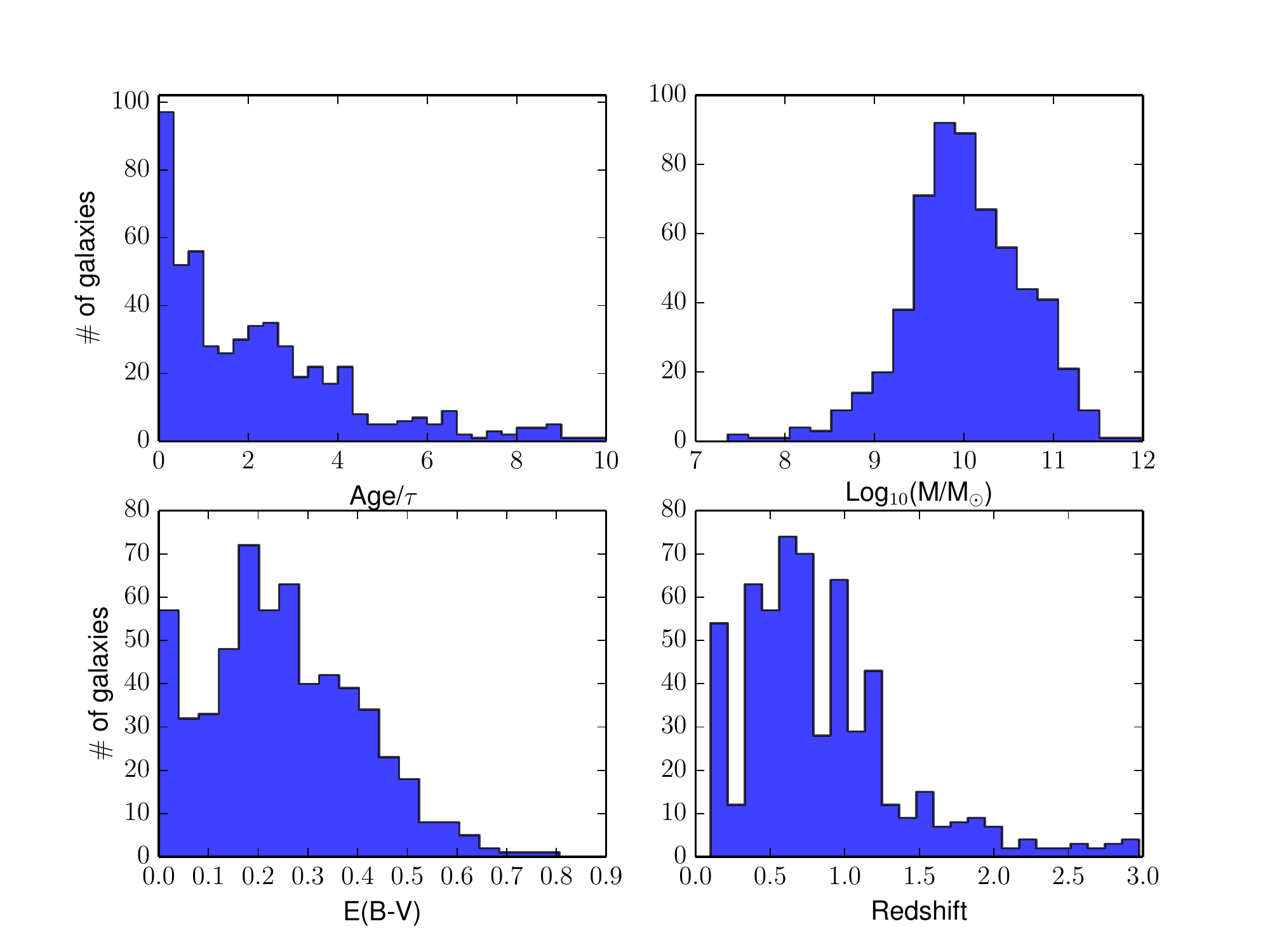}
\caption{Distribution of input parameters for the two mock catalogs characterized by constant SFH (left), and exponentially declining SFH (right).}
\label{Fig:input}
\end{centering}
\end{figure*}

One important difference between photo-z estimation and SED fitting is that if spectroscopic redshifts are available at least for a subset of the data, it is possible to test the performance of the photo-z algorithm, and this information can be used to iteratively improve the redshift estimation method. This luxury is usually not available for SED fitting parameters, since the ``true" values are unknown. A notable exception are the (few) galaxies for which Lick indices or spectroscopic line widths or ratios are available and provide an alternative estimate of age, mass, or metallicity. For mass or stellar age, tests conducted on mock catalogs will reveal systematic biases introduced by imperfect exploration of the parameter space (``algorithmic systematics"), but will be blind to systematic biases caused by, \eg catastrophic failures in the photometry or excessively simplistic SPS models. On the other hand, those causes of systematic uncertainties in parameter determination are often dominant with respect to those arising from uncertainties in the photometry (\eg \citealt{Conroy2013,Lee2009,Simha2014}; Mobasher et al. 2015).

In this paper, we show that the performance of photo-z estimation in a joint SED fitting+photo-z approach is a good tracer of these ``non-algorithmic systematics". We then demonstrate how these pieces of information can be used to improve the determination of the SED fitting parameters. We propose a technique to correct the estimation of uncertainties in redshift and other SED fitting parameters with minimal information loss, \ie without degrading the quality of parameter fitting.

The paper is organized as follows. We use mock data in Sections 2, 3 and 4. In Section 2 we show that the code we use, SpeedyMC, is able to recover redshifts and SED fitting parameters correctly in absence of ``non-algorithmic systematics". In Section 3 we investigate the separate and combined effects of adding catastrophic failures to the photometry, adding uncorrected zero-point errors, and of using incomplete libraries of SPS models. In Section 4 we  provide a strategy to recalibrate uncertainties in all parameters on the basis of photo-z information, and we estimate the effect of using empirical offsets, which are common in photo-z algorithms. In Section 5 we show that it is possible to apply the same strategy to data, for joint SED fitting and photo-z estimation, and for a two-step process where a photo-z algorithm is used before SED fitting. Section 6 summarizes our findings and our prescription for calibrating uncertainties in SED fitting parameters on the basis of uncertainties in photo-z.

\begin{figure*}
\begin{centering}
\includegraphics[width=0.49\linewidth]{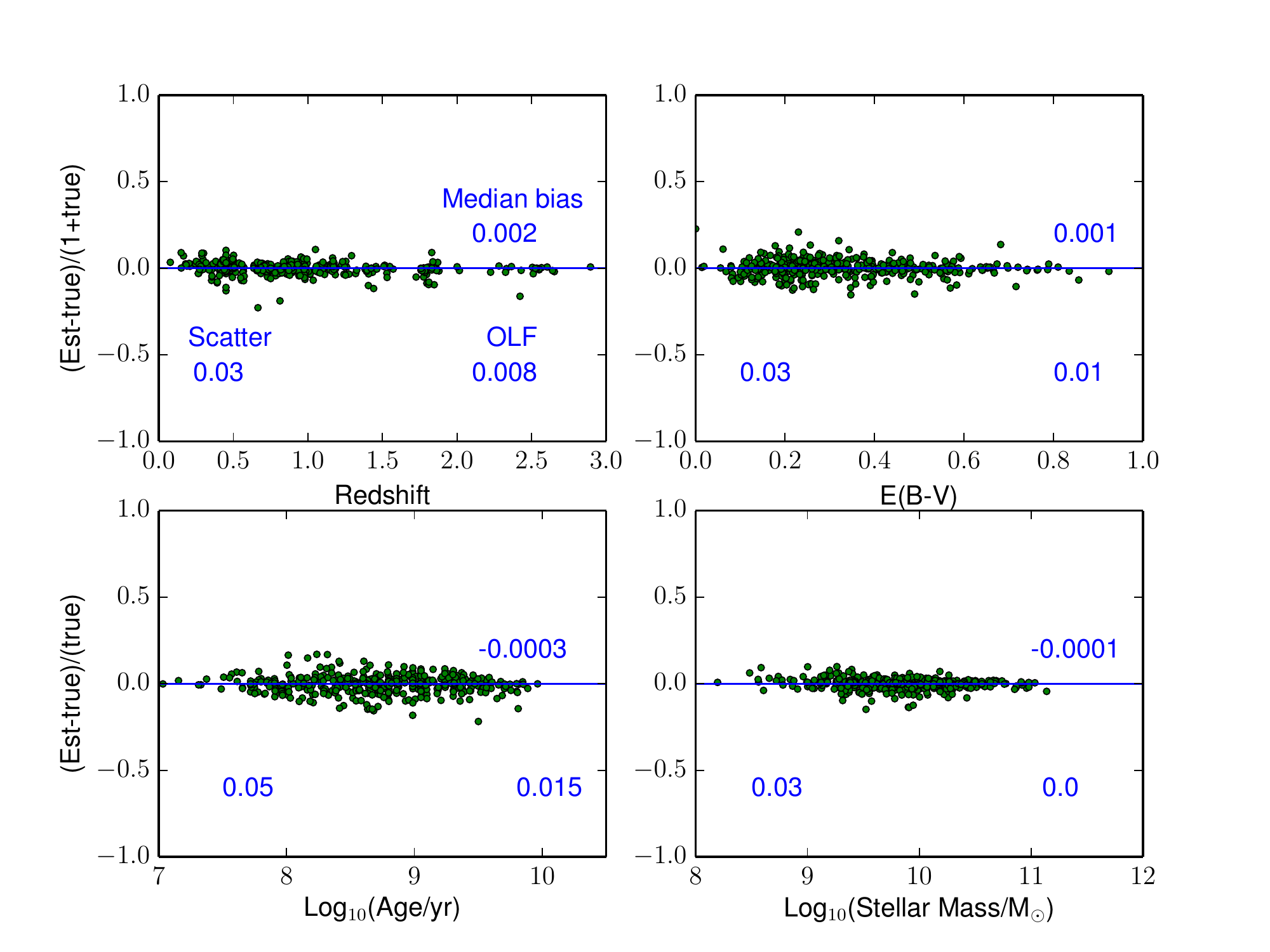}
\includegraphics[width=0.49\linewidth]{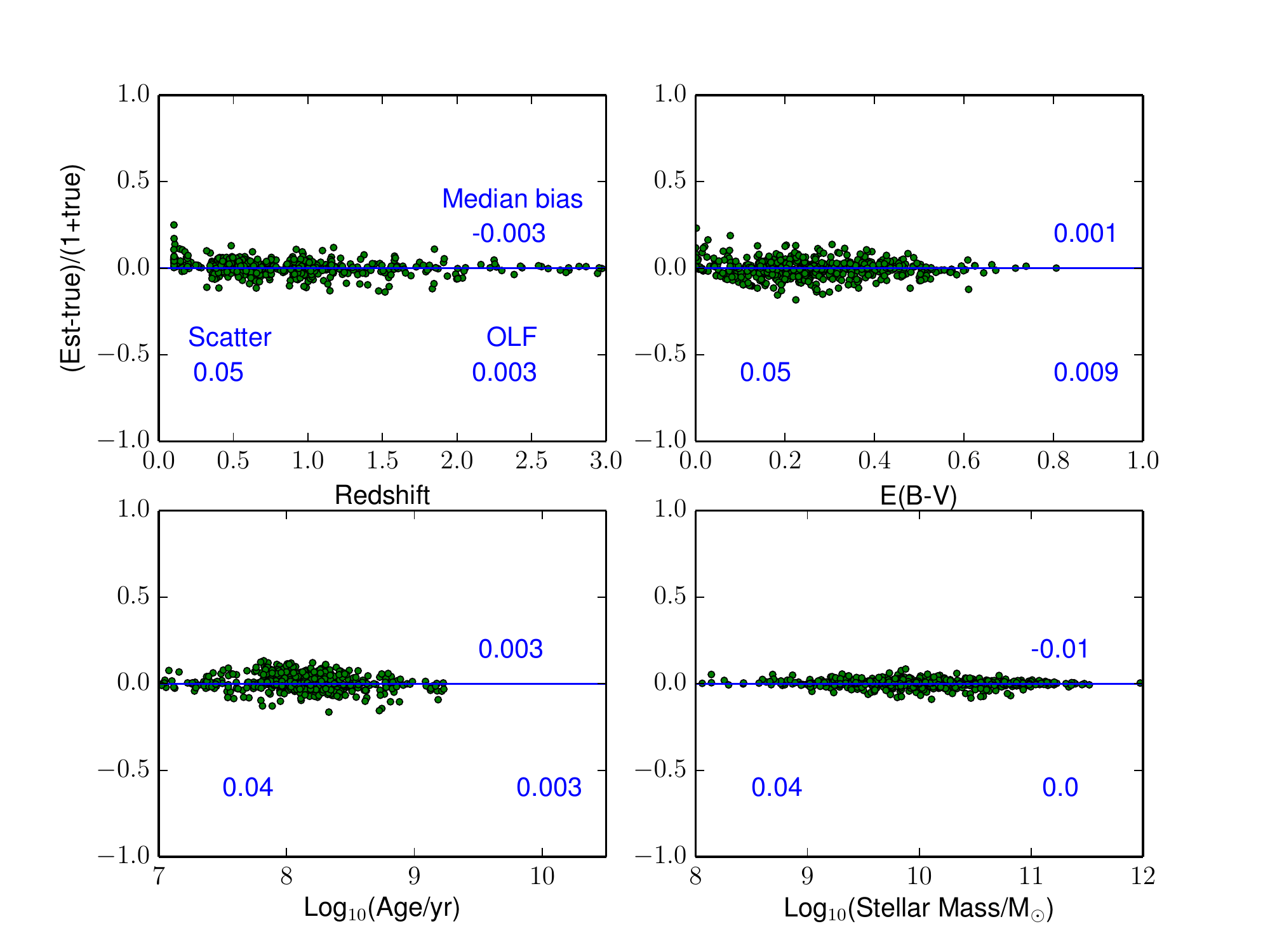}
\caption{Recovery of input parameters shown for several hundred mock galaxies generated with constant (left) and exponentially declining (right) star formation history. Noise and scatter are added to the photometry, but the mock data are fit using the correct functional form for the SFH and the photometry is immune from catastrophic errors. In absence of ``non-algorithmic systematics", SpeedyMC is able to correctly recover the input parameter values, as well as to correctly estimate the size of the uncertainties.} 
\label{fig:Mocks1}
\end{centering}
\end{figure*}

\begin{figure*}
\begin{centering}
\includegraphics[width=0.49\linewidth]{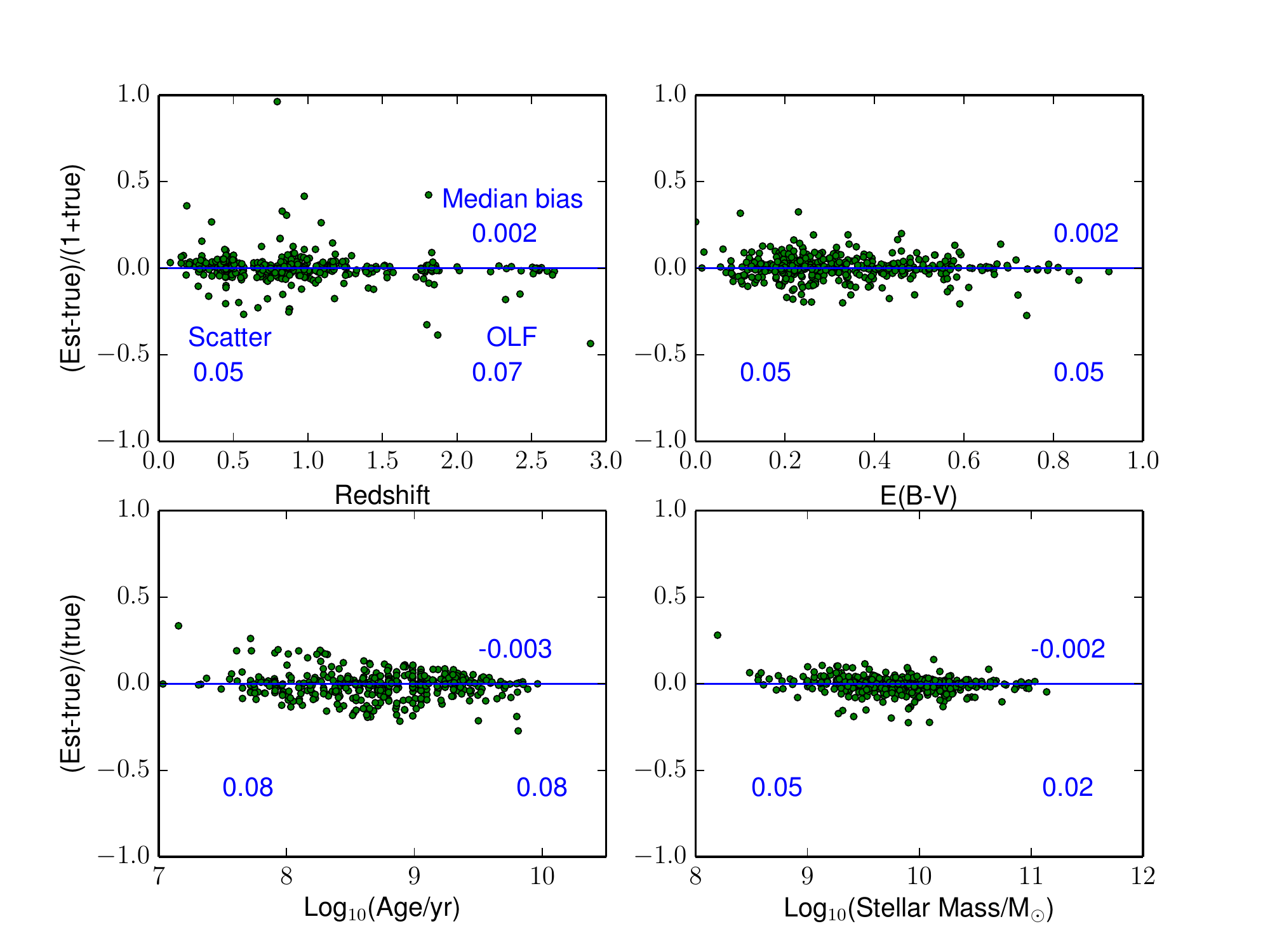}
\includegraphics[width=0.49\linewidth]{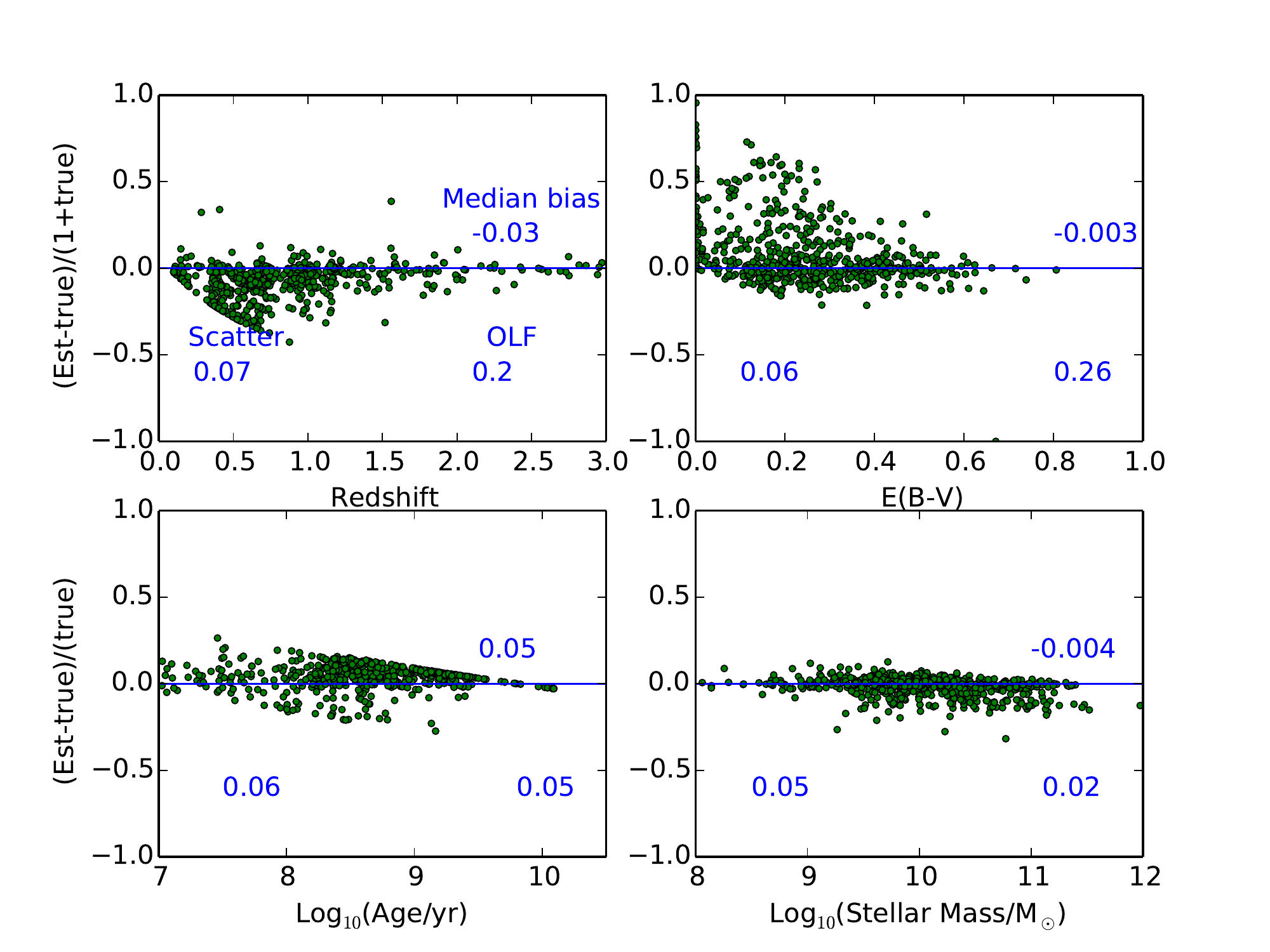}
\caption{This figure shows the effect of introducing catastrophic failures in the catalog generated with CSF in the form of a 10-$\sigma$ outlier in 3\% of data points, an example of a mild source of systematics, in the left panel, and of our ``worst case scenario", corresponding to using CSF to fit models with exponentially declining SFHs, combined with uncorrected zero-point errors and catastrophic failures, in the right panel.}
\label{fig:Mocks2}
\end{centering}
\end{figure*}

\section{Joint SED-photoz fitting with SpeedyMC}

We use the SpeedyMC code \citep{SpeedyMC} to jointly perform SED and photo-z fitting. SpeedyMC is a Markov Chain Monte Carlo code based on GalMC \citep{GalMC}. 
It uses a Metropolis-Hastings algorithm with adaptive proposal density to explore the parameter space, and it employs Bayesian statistics to calculate the posterior probability as a product of likelihood and priors. In the case of redshift, we adopt a luminosity function prior based on the evolving cosmological volume in a $\Lambda$CDM cosmology.

We create two mock catalogs, one characterized by constant star formation history (CSF), and one characterized by exponentially declining star formation history (ESF);  these two star formation histories are among the most common ones used in SED fitting. 

The model parameters are stellar mass, stellar age, dust content, parametrized by the excess color E(B-V), redshift, and e-folding time $\tau$, in the case of ESF. We define age as the median stellar age of the galaxy, \ie the lookback time in which 50\% of the stellar mass has been built. Whenever there is a mismatch between the ``true" and ``assumed" functional form for the star formation history, we compare two definitions of age: the median stellar age and the time elapsed since the onset of star formation. We demonstrate that in all cases the median stellar age is better recovered and is more robust to incorrect assumptions in the star formation of history of the galaxy. We adopt uniform priors in log(Age), log(Mass), log($\tau$), and E(B-V). 

For both sets of models, the input physical properties and broadband coverage are modeled after a bright subset of the Cosmic Assembly Near-Infrared Deep Extragalactic Survey (CANDELS, \citealt{Grogin11}) GOODS-S photometric catalog described in \cite{Guo2013}. The model galaxies are at $0 < z < 3$, and the UV-to-mid infrared multi-wavelength SED allows one to obtain deep coverage in the rest-frame UV to rest-frame NIR, which is essential for SED fitting. The distribution of input parameters for the two catalogs is shown in Figure \ref{Fig:input}; for ESF we show the ratio age/$\tau$ since it's more representative of the star formation mode of the galaxy.

Both catalogs comprise several hundred galaxies. We perturb the photometry in each band by a random Gaussian scatter of amplitude equal to 10\% of the flux in that band. We also verified that varying the fractional errors across the range of photometric bands does not affect our results. In particular, the ranking of the severity of systematics (with uncorrected zero point errors being the mildest, and template incompleteness being the most severe) is unaltered, and the degree of correlation between the underestimation of uncertainties in different parameters is unchanged. Additionally, an example application with a fully realistic distribution of photometric uncertainties is explored in Section \ref{Sec:Data}. 

To evaluate the performance of the code, we single out a few quantities of interest. The first is the difference between input parameter and output parameter. SpeedyMC outputs full marginalized probability distribution functions (PDFs) for each parameter, so one could consider either the best-fit value or the mean of the PDF. We chose to use the best-fit value since the mean might not be meaningful in the case of bimodal distributions, although in this case it doesn't make any appreciable difference. For redshift and E(B-V), whose values can be close to zero, we evaluate the difference (referred to as ``bias") as (estimated-true)/(1+true), while for Log(age) and Log(mass) we use the percent difference: (estimated-true)/(true). The ``scatter" is the standard deviation of these vectors and it represents the overall quality of the SED fitting. We are also interested in whether the uncertainties on the different parameters are estimated correctly. The ``true" uncertainties can be calculated by counting for how many objects the true (input) value for each parameter lies within the X\% estimated credible interval; for example, the width of the $68\%$ credible interval is correctly estimated if the true value is within this interval for 68\% of the objects, and under(over)-estimated if this happens for fewer (more) objects. Therefore, we can check whether the uncertainties were under- or over-estimated in each parameter by reporting the fraction of objects whose ``true" values lie within the estimated 68, 95\% credible regions.
Finally, we report the outlier fraction (OLF) for each parameter. For redshift and E(B-V), following \cite{Dahlen2013}, we define it as the fraction of objects for which (estimated-true)/(1+true) $>$ 0.15; for age, mass, and $\tau$ we consider as outliers those objects for which (estimated-true)/(true) exceeds 0.15. These definitions are based on the sampling variables (\ie $\log{(\rm Age/yr)}, \log{(\rm M/M_\odot)}$).  

The results for both sets of models are very encouraging. Figures \ref{fig:Mocks1} and \ref{fig:Mocks2} show that the median bias in each parameter is small (less than 0.005 in all cases), with the only exceptions of $\tau$, which isn't well constrained and for which the probability distribution functions of many objects are effectively flat. However, poorly constrained values of $\tau$ do not create any significant bias in the recovery of the other parameters. We experimented using different priors in $\tau$ and concluded that a uniform prior in log($\tau$) causes the least bias in other parameters. Also, the size of the 68 and 95\% credible regions are close to optimal for all parameters (other than $\tau$, for which the amplitude of the 68\% credible region is overestimated) in both cases. We note that the definition of outliers, which is arbitrary, would cause changes of the order of a few percent in the amplitude of the credible regions, setting the empirical precision at which the confidence interval can be estimated. Finally, the fraction of outliers is very small (within 1.5\%), again for all parameters other than $\tau$, in both cases. We conclude that, in absence of ``non-algorithmic systematics" such as template incompleteness and catastrophic errors in the photometry, SpeedyMC recovers the input parameters and estimates the size of the uncertainties correctly.

\section{SED fitting in presence of non-algorithmic systematics}
\label{Sec:Syst}

In this section we analyze the effect of three common sources of systematic bias: catastrophic failures in the photometry (i.e., photometric measurements that are far off the ``true" value), uncorrected zero-point errors in the photometry, and the use of incomplete or inadequate libraries of stellar population templates. The results for all these tests are summarized in Table \ref{Tab:syst}. \\

{\bf Effect of catastrophic failures.} We use the same set of mock galaxies with constant star formation history considered above, and revise the photometry so that every photometric point has a 3\% probability of being replaced by a 10-sigma outlier. The results of this test are shown in the left panel of Figure \ref{fig:Mocks2}. With respect to the systematics-free case (left panel of Figure \ref{fig:Mocks1}), we note that the best fit values now suffer from a (very) slight bias and the size of uncertainties in all parameters are moderately underestimated (by about 15\%). The fraction of outliers in redshift, age and E(B-V) increases significantly, from 1.5\% or less to 6.7, 7.5 and 5.0\% respectively. Finally, the overall scatter of true versus estimated parameters (in other words, the size of the uncertainties in the SED fitting parameters), which summarizes of how well parameters can be measured, increases by 20-30\% in each parameter. \\

{\bf Effect of uncorrected zero-point errors.} For the mock catalog with constant star formation history, we introduce a band-dependent systematic bias in order to mimic the effect of uncorrected zero-point errors in the photometry. We assume that the amplitude of this effect is 2\% of the flux in UV, optical, and NIR bands, 5\% in K-band, and 8\% in IRAC, as suggested by \eg the analysis in \cite{Guo2013}. The sign of the bias is random, but consistent within filters belonging to the same instrument. The results of this test can be found in Table \ref{Tab:syst}. The best fit values suffer from a slight bias (between 1 and 3\%) and the size of uncertainties in all parameters are underestimated by 10-15\%. The fraction of outliers in redshift, age and E(B-V) is low, 1, 5 and 2.5\% respectively, and negligible in mass. The overall scatter of true versus estimated parameters again increases moderately, by 20-30\%.  \\

{\bf Effect of using an incorrect Star Formation History.} Most SPS models used in SED fitting have relatively simple star formation histories. To capture the effect of this simplifying assumption on the parameter estimation, we generate models with a given star formation history (SFH), and then assume a different, incorrect SFH when performing SED fitting. We run this test for two sets of mock galaxies: for models generated with CSF and fit assuming a linearly increasing star formation rate ($\Psi(t) \propto t$), and for models generated with ESF and fit by assuming a constant star formation rate. We expect the bias introduced by our wrong assumptions to be mild in the first case, and severe in the second case.

The results of these tests confirm our expectations. Assuming a linearly increasing star formation history (when the correct one is constant) does not impact the correct recovery of mean values and credible regions for photo-z, stellar mass and E(B-V). If age is defined as the time elapsed from the beginning of star formation, the incorrect assumption on the star formation history induces a moderate but significant bias in this parameter (3\%, versus a scatter of 5\%), and increased scatter. The fraction of objects for which the input age values lie within the 68 and 95\% regions is significantly smaller than 68 and 95\%, as a result of the systematic bias. Conversely, the median stellar age is less sensitive to the mismatch between true and assumed star formation history, and the estimation of mean values and credible regions for median stellar age remains correct.

Fitting models whose SFH is exponential using a CSF has more drastic consequences. All parameters are affected and the underestimation of the uncertainties is quite dramatic (around 50\%). Redshift and mass are the parameters for which the effect is largest. The outlier fractions in redshift and E(B-V) are as high as 15\% and 24\% respectively. The bias and scatter found in the distribution of estimated-vs-true parameters are also higher with respect to the case in which ESF models are fit using the correct star formation history, although numbers change moderately because these calculations are made excluding outliers. Once again, the median stellar age is recovered from SED fitting much better than the time elapsed since the onset of star formation.  \\

{\bf Combined effect of several sources of modeling error in realistic simulations.} While the previous cases aimed at estimating the effect of individual sources of systematic errors, when using data the most common scenario is that several simplifying hypotheses are used at once. We attempt to simulate the effect of these combined factors by generating a new mock catalog, comprised of 1,000 galaxies, as follows. The star formation history of each galaxy is simulated as the combination of either an exponentially decreasing model (SFH $\propto e^{-\frac{t}{\tau}}$) or a delayed $\tau$ model  (SFH $\propto$ t$\, e^{\frac{-t}{\tau}}$), combined with a burst. The fraction of mass in the burst is a random variable uniformly distributed between zero and one, the age of the burst is randomly sampled from a uniform logarithmic distribution between 10 Myr and 1 Gyr. For the smooth SF component, the logarithms of ages and masses follow a uniform distribution (between 6.5 and 10 and between 8 and 12 respectively), and the ratio $\tau/age$ for the ESF or DSF model is randomly selected within the [0.1-10] interval. We also consider three possible (randomly selected) dust laws for the galaxies: the Calzetti law \citep{Calzetti2000}, the Milky Way dust law from \cite{Cardelli1989}, and the two-component model of \cite{Charlot2000}, where the birth cloud parameters $\tau_{BC}$ and $\mu_{BC}$ follow a Gaussian distribution with mean values 1.0 and 0.3 and standard deviations equal to 0.2 and 0.1, respectively. 

To include a realistic model of photometric noise, we use the CANDELS GOODS-S photometric catalog, fit the distribution of photometric uncertainties as a function of flux density in each band, and apply it to our mock catalog. We perform SED fitting on these catalogs using a single stellar population model with exponentially declining star formation history, and assuming that the dust attenuation law is the Calzetti law with the variable parameter E(B-V). To show the effect of using a single stellar population and an incorrect dust law separately, we split the two cases into the full catalog and a subset where the Calzetti law is used both in simulations and fitting (dubbed ``correct dust law" in Table \ref{Tab:syst}).

For the case where the incorrect functional form for the dust law is used, we do not report a significant bias for any of our SED fitting parameters other than $\tau$, which is poorly constrained. However, we observe a significant fraction of outliers (about one in eight objects) for all parameters. The uncertainties in all parameters are significantly underestimated, by a factor of 30-40\%. In this case where multiple stellar populations are present, it is especially preferable to use median stellar age to define the age of the galaxy; using time since the beginning of star formation would lead to a bias eight times greater and an even stronger underestimation of the uncertainties.

These unsatisfactory behaviors (underestimation of uncertainties, high OLF) are mitigated significantly when the correct functional form for the dust absorption law is used, with OLFs decreasing by 30-40\% and milder underestimation of the uncertainties in all other parameters.

We also attempted to model another source uncertainty in stellar evolution, the contribution of TP-AGB stars, by using the stellar population models from Bruzual and Charlot (2007) to fit the mock catalogs obtained with the 2003 stellar population synthesis models. We note that this issue (as well as the well-known uncertainty in the IMF) leads to a systematic shift in the estimate stellar masses rather than to a true underestimation of the uncertainties related to a certain measurement, and therefore we haven't included it in our prescription for calculating the annealing temperature. Using the CB07 models to fit a CB03 ``truth" causes an increase in the fraction of outliers in redshift of a factor of 2 (to a total of 26\%), producing a notable ``red flag" in the results of the SED fitting. \\

{\bf Effect of combined systematics.} Our final scenario is the ``worst-case" one in which catastrophic failures in the photometry, uncorrected zero point errors, and template incompleteness all affect the SED fitting. Unfortunately, unless our treatment of the systematics is pessimistic, this is also the most realistic scenario when using data, and therefore particularly worthy of our consideration. As anticipated and shown in the last row of Table \ref{Tab:syst}, the combined effect of these three systematics makes the results even worse than the previous case, with outlier fractions in redshift and E(B-V) reaching 21\% and 29\% respectively, and even more severely underestimated uncertainties in all cases. This case is illustrated in the right panel of Figure \ref{fig:Mocks2}.

\begin{table}
\begin{center}
\resizebox{\linewidth}{!}{
\begin{tabular}{|c|cccccc|}
\hline
& Parameter & \% of objects & \% of objects & median & scatter & OLF  \\
& & in 68\% region & in 95\% region & bias  & & \\ 
 \hline
 & z & 67 & 94 & 0.002 & 0.03 & 0.008 \\
CSF & Mass  & 69 & 93 & -0.0001 & 0.03 & 0.0 \\
 & Age (median)& 68 & 94 & -0.0003 & 0.05 & 0.015 \\
& E(B-V) & 67 & 94 & 0.001 & 0.04 & 0.01 \\
 \hline
& z & 57 & 84 & 0.0008 & 0.04 & 0.07 \\
CSF & Mass  & 60 & 85 & -0.001 & 0.04 & 0.02 \\
+ Cat Fail & Age (median) & 62 & 87 & -0.003 & 0.06 & 0.08 \\
& E(B-V) & 59 & 87 & 0.002 & 0.05 & 0.05 \\
\hline
& z & 55 & 90 & -0.01 & 0.03 & 0.01 \\
CSF & Mass  & 59 & 90 & 0.01 & 0.04 & 0.0 \\
+ ZP errors & Age (median) & 56 & 91 & 0.03 & 0.06 & 0.05 \\
& E(B-V) & 59 & 90 & -0.02 & 0.05 & 0.03 \\
\hline
& z & 67 & 94 & -0.0002 & 0.03 & 0.008 \\
CSF & Mass  & 68 & 90 & 0.002 & 0.03 & 0.0 \\
fit as Lin Inc & Age (onset) & 50 & 88 & 0.03 & 0.05 & 0.02 \\
& Age (median) & 68 & 90 & 0.007 & 0.05 & 0.015 \\
& E(B-V) & 65 & 88 & 0.003 & 0.04 & 0.01 \\
\hline
& z & 61 & 91 & -0.003 & 0.05 & 0.003 \\
& Mass  & 68 & 96 & -0.01 & 0.04 & 0.0 \\
ESF & Age (median) & 68 & 94 & 0.003 & 0.04 & 0.005 \\
& E(B-V) & 66 & 93 & 0.001 & 0.05 & 0.009 \\
& $\tau$ & 84 & 94 & 0.006 & 0.2 & 0.25 \\
\hline
& z & 38 & 67 & -0.02 & 0.05 & 0.15 \\
ESF & Mass  & 51 & 78 & -0.01 & 0.04 & 0.01 \\
fit as CSF & Age (onset) & 48 & 82 & 0.03 & 0.05 & 0.01 \\
 & Age (median) & 60 & 88 & 0.02 & 0.05 & 0.007 \\
& E(B-V) & 58 & 81 & 0.01 & 0.05 & 0.24 \\
\hline
& z & 38 & 68 & -0.02 & 0.05 & 0.18 \\
ESF & Mass  & 46 & 72 & -0.01 & 0.04 & 0.02 \\
fit as CSF & Age (onset) & 49 & 71 & 0.02 & 0.06 & 0.06 \\
+ Cat Fail & Age (median) & 56 & 80 & 0.015 & 0.05 & 0.04 \\
& E(B-V) & 54 & 76 & 0.02 & 0.05 & 0.24 \\
\hline
& z & 54 & 86 & -0.01 & 0.05 & 0.07 \\
ESF & Mass & 56 & 84 & 0.006 & 0.04 & 0.01 \\
+ Cat Fail & Age (median) & 55 & 85 & 0.02 & 0.05 & 0.04 \\
+ ZP errors & E(B-V) & 59 & 88 & -0.02 & 0.06 & 0.11 \\
& $\tau$ & 83 & 92 & 0.13 & 0.3 & 0.38 \\

\hline
& z & 46 & 67 & -0.004 & 0.04 & 0.125 \\
Realistic SFH & Mass & 38 & 61 & -0.005 & 0.03 & 0.01 \\
fit as single-pop; & Age (onset) & 27 & 45 & -0.03 & 0.06 & 0.13 \\
incorrect dust law & Age (median) & 40 & 60 & -0.004 & 0.05 & 0.12 \\
 & E(B-V) & - & - & - & - & - \\
& $\tau$ & 35 & 46 & 0.05 & 0.21 & 0.39 \\

\hline
& z & 30 & 47 & 0.004 & 0.08 & 0.26 \\
Realistic SFH & Mass & 13 & 19 & -0.08 & 0.05 & 0.58 \\
fit as single-pop; & Age (onset) & 26 & 40 & -0.02 & 0.06 & 0.12 \\ 
with CB07 models & Age (median) & 30 & 44 & -0.03 & 0.06 & 0.13 \\
\& incorrect dust law & E(B-V) & - & - & - & - & - \\
& $\tau$ & 33 & 43 & 0.07 & 0.24 & 0.34 \\

\hline
& z & 61 & 80 & 0.002 & 0.03 & 0.08 \\
Realistic SFH & Mass & 53 & 76 & -0.002 & 0.02 & 0.003 \\
fit as single-pop; & Age (onset) & 29 & 49 & -0.03 & 0.05 & 0.1 \\ 
correct dust law & Age (median) & 48 & 69 & -0.003 & 0.05 & 0.09 \\
& E(B-V) & 62 & 81 & 0.003 & 0.03 & 0.025 \\
& $\tau$ & 34 & 44 & 0.08 & 0.19 & 0.38 \\

\hline
& z & 32 & 57 & -0.03 & 0.07 & 0.2 \\
ESF & Mass & 43 & 65 & -0.004 & 0.05 & 0.02 \\
fit as CSF & Age (onset) & 34 & 67 & 0.06 & 0.06 & 0.08 \\
+ Cat Fail + ZP errors  & Age (median) & 39 & 74 & 0.05 & 0.06 & 0.05 \\
& E(B-V) & 47 & 74 & -0.003 & 0.06 & 0.26 \\
\hline
\end{tabular}}
\caption{Results for different tests on mock catalogs whose photometry is affected by ``non-algorithmic systematics" (catastrophic failures in the photometry, template incompleteness, uncorrected zero point errors, or all of them). When there is an ambiguity in the definition of age, we report results using both the median stellar age and the age since the onset of star formation, to show that the former is recovered better in all cases. When flexible dust laws are used, the input parameter E(B-V) is not well defined.}
\label{Tab:syst}
\end{center}
\end{table}

\begin{figure}
\begin{centering}
\includegraphics[width=\linewidth]{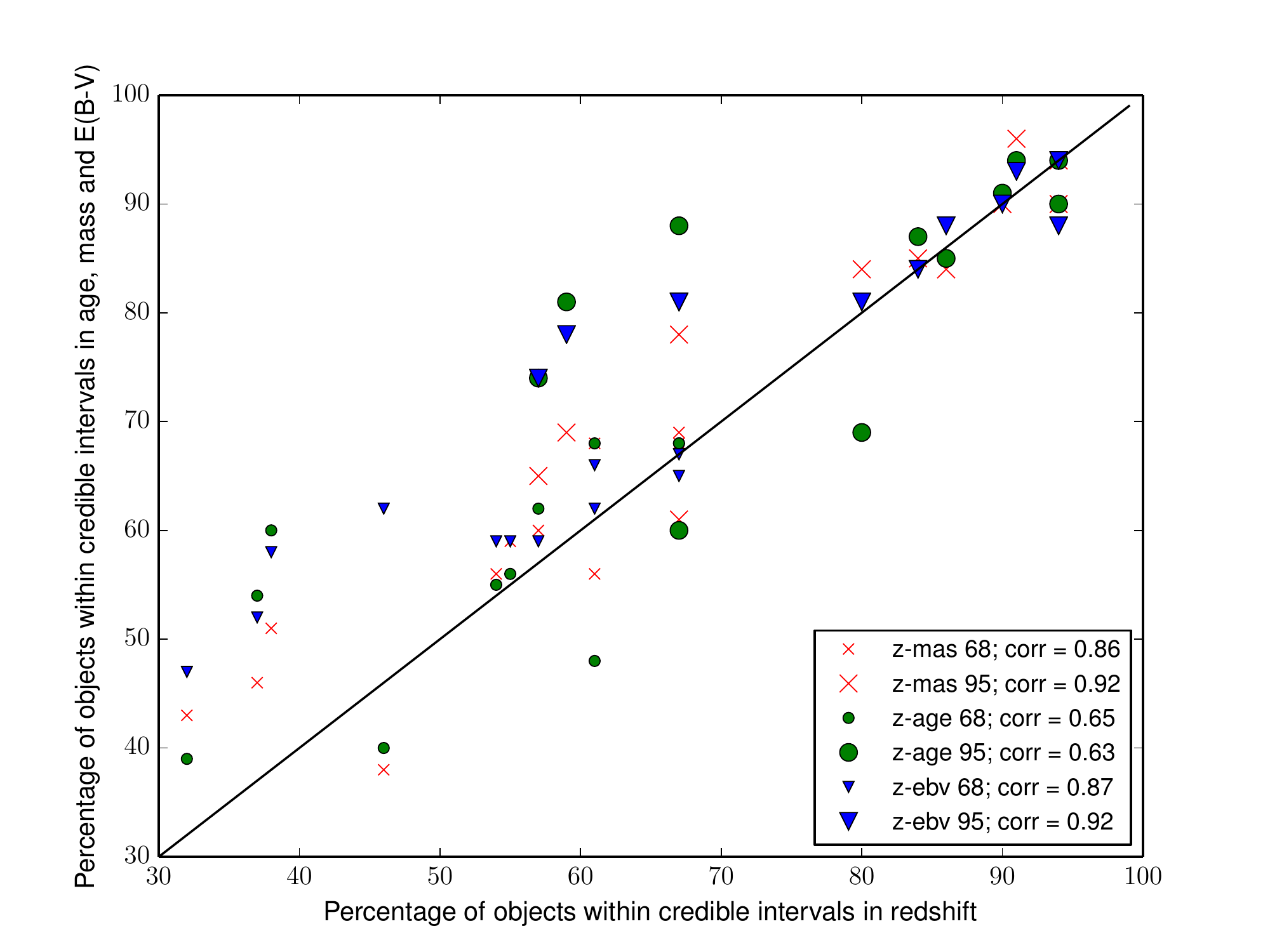}
\caption{Percentage of objects within the estimated credible intervals for all the example cases considered in Section \ref{Sec:Syst}. The legend shows Pearson correlation coefficients of the values obtained for redshift (x-axis) and other SED fitting parameters. A perfect correlation would indicate that if the uncertainties in redshift are underestimated by a certain amount, those in mass, age, or E(B-V) are underestimated by the same amount. Small (large) symbols refer to 68 (95)\% credible intervals.} 
\label{fig:correlation}
\end{centering}
\end{figure}

\section{Improved estimation of uncertainties on SED fitting parameters}
\label{Sec:improv}

The results from the tests on mock galaxies described in the previous sections (and the parallel ones we conducted with different prescription for simulated noise, which confirm these findings without adding new information) teach us some lessons about the effect of ``non-algorithmic" systematics on SED fitting. In particular, we note that:
\begin{itemize}
\item The fraction of outliers in redshift is a meaningful ``red flag" for the presence and gravity of these systematics. In our tests, the fraction of outliers rose from less than 1.5\% in the case of no catastrophic failures and correct SPS templates, to a few percent when catastrophic failures were added, to over 20 percent in the case of combined catastrophic failures, uncorrected zero-point errors and incorrect SPS templates. If the expected levels of catastrophic photometric failures and zero-point errors are known, the expected OLF from those can be calculated from simulations, and one could use any discrepancy to evaluate, and possibly correct, residual systematics deriving from template incompleteness.
\item Whenever non-algorithmic systematics are present, the size of uncertainties in photometric redshift as well as in SED fitting parameters is underestimated. The extent of this underestimation can be severe even in very common circumstances. Therefore, it is important to find a way to test and correct the size of uncertainties. 
\item The size of estimated uncertainties in redshift  is {\it strongly} correlated with the size of estimated uncertainties in other parameters, as shown in Figure \ref{fig:correlation}, even in critical cases of several incorrect assumptions made in SED fitting.
For the eleven tests shown in Table \ref{Tab:syst}, we calculated the Pearson correlation coefficient between the percentage of objects within the 68 and 95\% credible intervals in redshift, and those of age, stellar mass, and E(B-V). The correlation is highest for mass (0.86 and 0.92 for the 68 and 95\% intervals respectively) and E(B-V) (0.87 and 0.92), and strong for age (.65 and .63). In other words, if the size of uncertainties in redshift is correctly estimated, so are the ones in SED fitting parameters; and if they are underestimated in redshift, they will be underestimated in the other parameters, especially stellar mass, by approximately the same amount. This is a useful piece of information because the relation between estimated and true size of uncertainties in redshift can be measured if spectroscopic redshifts are available for a subset of galaxies. As a result, we argue that it is possible to use the estimated-vs-true size of uncertainties in redshift to {\it approximately calibrate} the size of uncertainties in other parameters, even in the presence of non-algorithmic systematics.
\end{itemize}

\noindent
Based on these results, we seek a method to simultaneously correct the size of uncertainties in photo-zs and SED fitting parameters. We assume that the factor by which uncertainties are mis-estimated in redshift can be obtained by testing the performance of the photo-zs against a control sample with spectroscopic redshifts. 

A first step in this direction is to estimate the unknown systematic errors by investigating the distribution of residuals between templates and observed photometry in each band. The width of this distribution provides a band-dependent smoothing error that can be added in quadrature to the photometric errors (\eg Dahlen et al 2013). However, as further discussed in Section \ref{Sec:Data}, there is no guarantee that this procedure will bring the uncertainties to a realistic level. In this case, a prescription is needed to modify uncertainties so that 1. the size of predicted and true uncertainties in redshift match (\ie the reported values lie within the N\% confidence errors N\% of the time); and, 2. there is no information loss in terms of bias, scatter, and OLF.
Our question then becomes: is it possible to modify the multi-dimensional PDF itself in a way that doesn't affect the relative weight of each band, and yet achieves the goal of re-sizing the uncertainties? Our proposed answer is to apply a technique called ``MCMC sampling at high temperatures" or, more often, {\it annealing}.

\subsection{Annealing}
\label{Subsec:ann}
Annealing (\eg \citealt{mackay}) is a technique used in MCMC sampling to speed up convergence in multi-modal probability distributions. In MCMC parlance, this process is described as increasing the ``temperature" T of the chains; the standard temperature is T = 1. This is achieved by replacing the likelihood of each model, $\cal{L}$, with ${\cal L}^{(1/T)}$. The effect of annealing is to increase the likelihood of improbable transitions. As a result, points occupying the tail of a probability distribution are sampled more often than they would be at low temperature. Because in MCMC the density of visited points is proportional to the probability distribution function, for single-peak distributions the PDF inferred from high-temperature chains has a similar mean and location of maximum as its low-temperature equivalent, but it has wider tails. In presence of multiple local minima, the secondary ones might get ``boosted" to higher significance. The practical effect on the marginalized distributions is to increase the size of the credible regions, achieving our objective of re-sizing the uncertainties of photo-zs as well as SED fitting parameters, but without information loss in the form of increased bias or a significantly higher fraction of outliers.

\begin{figure*}
\begin{centering}
\includegraphics[width=0.49\linewidth]{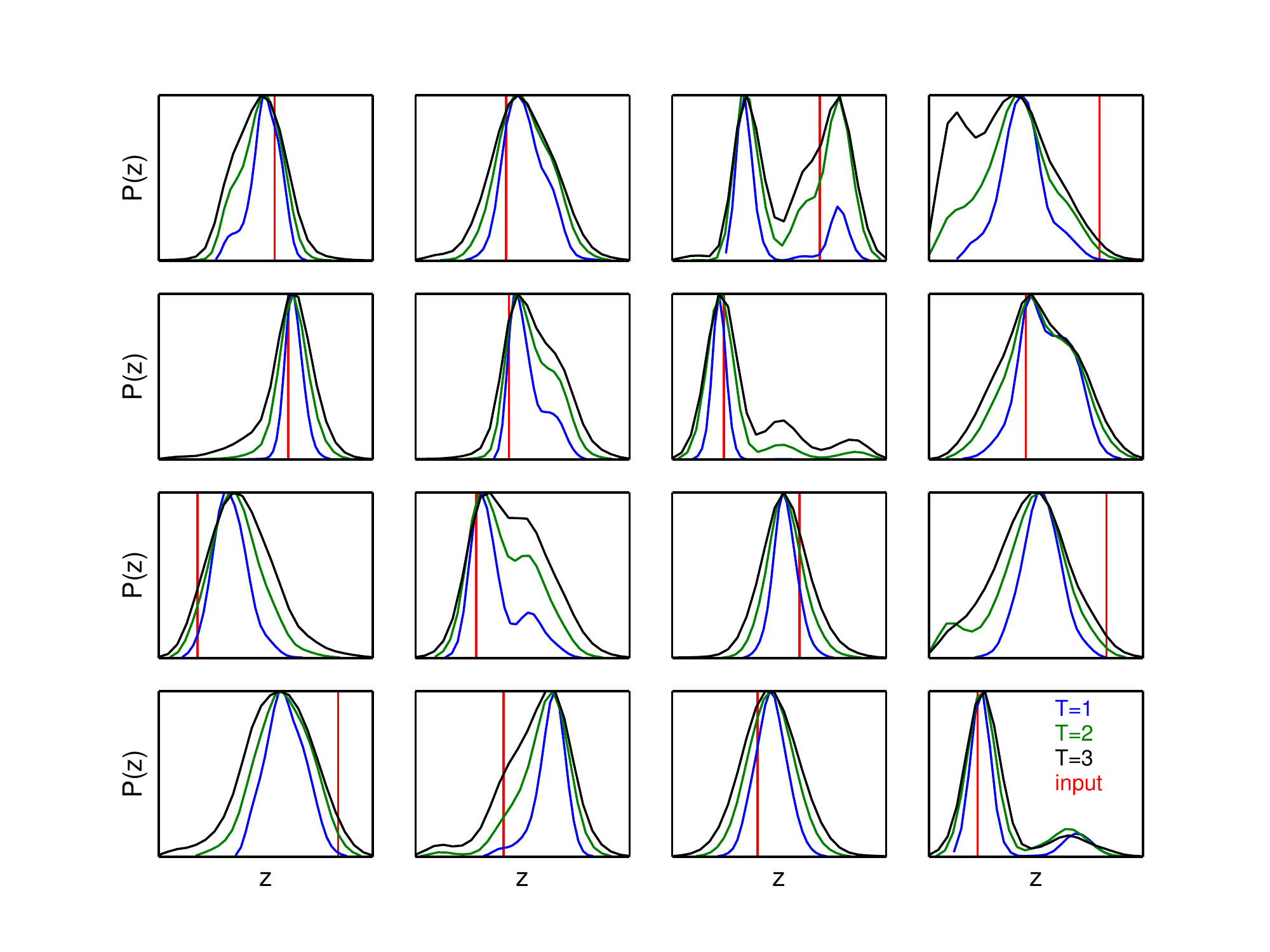}
\includegraphics[width=0.49\linewidth]{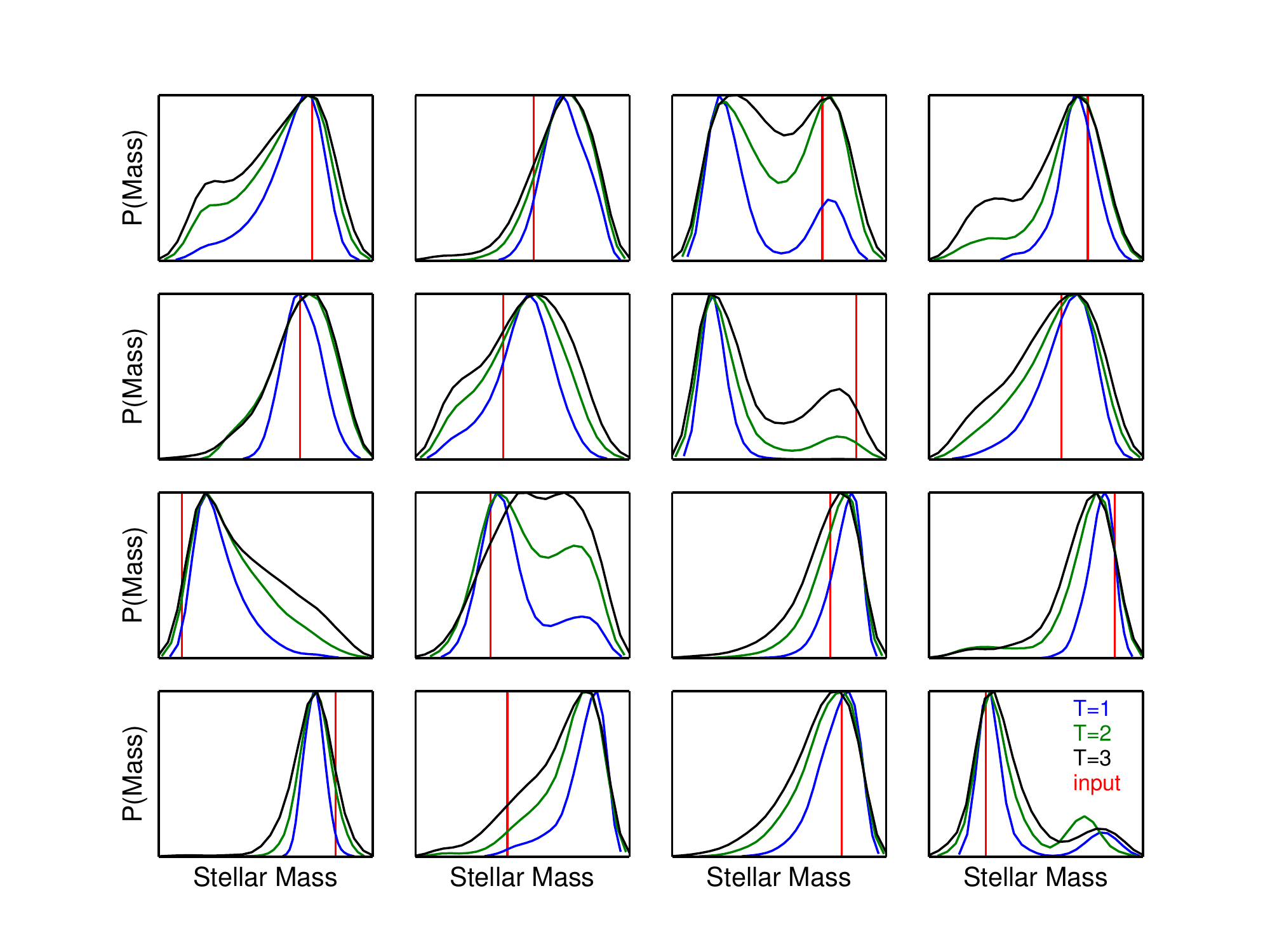}
\caption{Effect of annealing, shown for the first 16 objects of our mock catalog generated with constant star formation history and added catastrophic failures in the photometry. The blue (narrowest) curves come from sampling at T = 1, the green curves come from sampling at T = 2, the black (widest) curves come from sampling at T = 3, and the red vertical lines show the true (input) values for redshift (left) and stellar mass (right). In this case, T = 2 is close to optimal.}
\label{fig:Annealing}
\end{centering}
\end{figure*}

By sampling the chains for the tests described above at different temperatures, we were able to verify that the width of the PDF distribution increased, but the bias and OLF did not become significantly worse when chains were run at higher temperature. For example, for the case of CSF with catastrophic failures, the fraction of objects for which the true value of the redshift is within the 68 (95) \% estimated credible region is 56 (83) \% at T = 1, becomes 68 (90)\% at T = 2, and 75 (94)\% at T = 3. More importantly for our purpose, the correlation between the size of credible regions in redshift and other SED fitting parameters holds. For example, the fraction of objects for which the true value of the mass is within the 68\% (95)\% estimated credible region varies from 59 (84)\% at T = 1, to 70 (90)\% at T = 2, to 73 (93)\% at T = 3. In each of three cases, the bias and OLF do not change significantly. An example of the effect of sampling at higher temperature on the recovered marginalized PDFs in redshift and mass is shown in Figure \ref{fig:Annealing} for the first 16 objects in our mock catalog with constant SFH and added catastrophic failures in the photometry.

\begin{table}
\begin{center}
\resizebox{\linewidth}{!}{
\begin{tabular}{|c|cccccc|}
\hline
& Parameter & \% of objects & \% of objects & median & scatter & OLF  \\
& & in 68\% region & in 95\% region & bias  & & \\ 
 \hline
 & z & 67 (57) & 89 (84) & 0.003 (0.002) & 0.07 (0.05) & 0.07 (0.07) \\
CSF& Age (median) & 72 (62) & 91 (87) & -0.003 (-0.003) & 0.06 (0.06) & 0.08 (0.08) \\
+ Cat Fail & Mass  & 69 (60) & 90 (85) & -0.002 (-0.002) & 0.04 (0.04) & 0.02 (0.02) \\
T = 1.5 & E(B-V) & 69 (59) & 91 (87) & 0.003 (0.002) & 0.05 (0.05) & 0.05 (0.05) \\
\hline
& z & 67 (38) & 91 (67) & -0.03 (-0.02) & 0.06 (.05) & 0.15 (0.15) \\
ESF & Age (median) & 69 (60) & 87 (88) & 0.02 (0.02) & 0.05 (0.05) & 0.01 (0.01) \\
fit as CSF & Mass  & 65 (51) & 84 (78) & -0.009 (-0.01) & 0.04 (0.04) & 0.02 (0.01) \\
T = 5 & E(B-V) & 69 (58) & 83 (81) & 0.01 (0.01) & 0.05 (0.05) & 0.24 (0.24) \\
\hline
ESF & z & 62 (37) & 88 (59) & -0.03 (-0.02) & 0.09 (0.09) & 0.22 (0.21) \\
fit as CSF & Age (median) & 85 (54) & 95 (81) & 0.02 (0.03) & 0.06 (0.06) & 0.09 (0.08) \\
+ Cat Fail & Mass  & 63 (46) & 81 (69) & -0.01 (-0.01) & 0.05 (0.05) & 0.05 (0.04) \\
T = 5 & E(B-V) & 71 (52) & 89 (78) & 0.01 (0.01) & 0.06 (0.06) & 0.31 (0.29) \\
\hline
ESF & z & 58 (32) & 86 (57) & -0.04 (-0.03) & 0.08 (0.07) & 0.21 (0.20) \\
fit as CSF & Mass & 57 (43) & 79 (65) & -0.005 (-0.004) & 0.05 (0.05) & 0.03 (0.02) \\
+ Cat Fail + ZP errors & Age (median) & 86 (39) & 95 (74) & 0.05 (0.05) & 0.06 (0.06) & 0.06 (0.04) \\
T = 5 & E(B-V) & 69 (47) & 86 (74) & -0.006 (-0.003) & 0.06 (0.06) & 0.27 (0.26) \\
\hline
& z & 64 (46) & 87 (67) & -0.004 (-0.004) & 0.04 (0.04) & 0.124 (0.125) \\
Realistic SFH & Mass & 59 (38) & 84 (61) & -0.005 (-0.005) & 0.03 (0.03) & 0.02 (0.01) \\
fit as single-pop & Age (median) & 56 (40) & 79 (60) & -0.005 (-0.004) & 0.05 (0.05) & 0.13 (0.12) \\
+ incorrect dust law & E(B-V) & - & - & - & - & - \\
T = 5 & $\tau$ & 43 (35) & 51 (46) & 0.06 (0.05) & 0.21 (0.21) & 0.38 (0.39) \\
\hline
& z & 77 (61) & 92 (80) & 0.0026 (0.002) & 0.03 (0.03) & 0.09 (0.08) \\
Realistic SFH & Mass & 73 (56) & 90 (84) & -0.002 (0.006) & 0.03 (0.04) & 0.01 (0.01) \\
fit as single-pop & Age (median) & 60 (48) & 81 (69) & -0.002 (-0.003) & 0.05 (0.05) & 0.1 (0.1) \\
+ correct dust law & E(B-V) & 75 (62) & 93 (81) & 0.004 (0.003) & 0.03 (0.03) & 0.022 (0.025) \\
T = 5 & $\tau$ & 37 (34) & 47 (44) & 0.04 (0.08) & 0.22 (0.19) & 0.35 (0.38) \\

\hline
\end{tabular}}
\caption{Effect of applying annealing to several of our ``problematic" test cases, including catastrophic failures, template incompleteness, uncorrected zeropoint errors, and their combined effect. In all cases, the estimation of uncertainties can be improved by using the spectroscopic redshifts control sample as a calibration tool. The numbers in parentheses indicate the results obtained without annealing.} 
\label{Tab:Ann}
\end{center}
\end{table}

We ran MCMC chains using annealing for seven problematic cases described in Table \ref{Tab:syst}, and we show our results in Table \ref{Tab:Ann}.  In each instance (catastrophic failures, template incompleteness, incorrect assumptions in modeling the stellar populations, uncorrected zero-point errors, combined effect) annealing was useful in improving the accuracy of the reported error bars on photo-zs and SED fitting parameters, without compromising the performance of SED fitting (\ie, without introducing further biases or increasing the fraction of outliers). A rule-of-thumb to determine the correct temperature for MCMC sampling (\eg the temperature for which the uncertainties will be close to their nominal value) is the following. In the chains, the comparison between different models that decided whether to accept or reject a step is done by comparing ${\cal L}^{(1/T)}$, where $\cal{L}$ is the likelihood of each model and T is the temperature of the chains. If the likelihood function were represented by a Gaussian function, an N-fold increase in temperature would correspond to an increase of $\sqrt{N}$  in the width of the likelihood (and therefore, in the size of uncertainties). Because the likelihood function is not Gaussian, and because of the presence of priors, the actual increase factor is different, and finding the ideal temperature for annealing might require some further trial-and-error. 

Furthermore, since the annealing temperature is calibrated using subsets of objects with spectroscopic redshifts, which might be on average brighter than photometric catalogs, we investigated how S/N affects the annealing temperature. We divided our mock catalogs into the high- and low- S/N halves, we ran chains with the same annealing temperature, and we calculated the accuracy of uncertainties for the two subsets. We found a mild dependence of the uncertainties on signal-to-noise, with uncertainties tending to be more severely under-estimated (by about 20\%) for the high S/N objects than for low S/N objects. As a result, the calibration obtained through annealing on a spectroscopic sample can be considered as an {\it upper} limit to the true uncertainties.

It is worth noting that as non-algorithmic systematics become more severe, the correlation between under (over) estimation of error bars between different parameters deviates significantly from the 1:1 relationship. For example, as can be seen in the first and second example, if the sizes of 68 and 95\% credible intervals are calibrated through redshift (as it would be for real data since for other parameters the true values are unknown), the corresponding uncertainties in stellar mass will be correct, but those in age and E(B-V) might be overestimated. Another severe instance of this problem was observed when using an incorrect dust attenuation law (using the Calzetti law to fit models in which the dust attenuation was generated using the Milky Way law and/or the Charlot and Fall model).
Therefore, {\it one should still attempt to correct the non-algorithmic systematics}, ``flagged" by large numbers of outliers, by using more flexible SPS models and more realistic star formation histories, before applying annealing to the data.
 
\subsection{Photometry offsets}
\label{Subsec:offsets}
It is common practice in photometric redshift codes (or SED fitting codes that can handle redshift) to account for unknown systematics, from template incompleteness to uncorrected zero-point errors to inaccurate filter system response by applying offsets to the photometry (\eg \citealt{Hildebrandt2012, Schmidt2013}). These offsets are calculated on the basis of the difference between the photometry of the best-fitting template for each galaxy and the data, and have been shown to improve the redshift estimation (\eg \citealt{COSMOSz}). Here we investigate the effect of applying those offsets on other SED fitting parameters. To this end, we use the last case shown in Table \ref{Tab:syst}, ESF models including catastrophic failures and uncorrected zero-point errors, which have been erroneously fit assuming a CSF history, and we:
\begin{itemize}
\item Calculate the distribution of residuals (difference between the flux of the best-fit template and the data flux) in each observed band; 
\item Fit the residuals distribution to a Gaussian curve and calculate its mean value;
\item Correct the SEDs by applying these offsets to the observed SEDs;
\item Repeat these three steps iteratively until the offsets stop changing, and the median $\chi^2$ of the models stop improving.
 \end{itemize}
 In our case, the above procedure required four iterations. 

Applying the offsets improved the redshift determination as expected: the median bias changed from -0.03 to -0.01, the OLF improved slightly from 20\% to 17\%, and the fraction of objects for which the true values were within the estimated 68 and 95\% intervals improved from 32 and 57\% to 42 and 71\% respectively. However, in the case of other parameters we did not notice any improvement. In fact, the scatter in stellar age and stellar mass became slightly worse (by $\sim 10\%$), the fraction of true values lying within the estimated 68 and 95\% intervals was unchanged (and thus remained severely underestimated) for stellar mass and E(B-V), and the fraction of outliers in E(B-V) became slightly worse, from 26 to 28\%. While this example is certainly insufficient to describe what could happen in general and with data, it supports the general prescription of using more flexible SPS models and using the OLF as a tracer of the severity of systematics, rather than relying on empirically determined offsets in photometry.
 
\section{Example application to data}
\label{Sec:Data}

The last step of our analysis is to verify that using annealing on real data, rather than on simulated galaxies, is still helpful in calibrating the uncertainties and does not introduce additional bias and/or worsen the fraction of outliers. In this case, we can only test the effect of annealing on redshift, since it's the only parameter for which the ``true" value can be learned via spectroscopic observations.

In order to perform join SED and photo-z fitting on real data, we used the publicly available multi-wavelength catalog obtained by CANDELS in the GOODS-S field, described in \cite{Guo2013}. The spectroscopic sample from this catalog comprises 1338 galaxies; we selected a subsample of 585 objects characterized by a signal-to-noise ratio (S/N) of at least 8 in at least 13 bands from observed-frame UV to IR. To determine photometric redshift and other physical properties using SpeedyMC, we adopt the following pipeline:

\begin{itemize}

\item Perform photo-z estimation and SED fitting jointly, including a luminosity function prior that takes into account the cosmological volume as a function of redshift. The distribution of estimated redshifts will have bias and scatter.
\item Calculate the distribution of residuals in each observed band as described in the previous section;
\item Use the width of the residual distribution as an estimate of the smoothing error in each band, to be added in quadrature to the photometric error;
\item Increase the uncertainties in the data by incorporating the smoothing errors in the photometry;
\item Evaluate any mismatch of the calculated and true uncertainties in redshift (\ie calculate for how many objects the spectroscopic redshift value lies within the 68 and 95\% estimated credible intervals);
\item Run MCMC chains at higher temperature as appropriate.
\end{itemize}

For these data, using a constant star formation history leads to a sizable median bias, scatter, and outlier fraction of -0.025, 0.06 and 16\%. The sizes of the 68 and 95\% credible intervals are also quite severely underestimated, at 42 and 66\% respectively. We note that this result is obtained after applying the smoothing errors; those are insufficient to ensure that the uncertainties in the SED fitting parameters will be estimated correctly, as mentioned in Section \ref{Sec:improv}. The high OLF alone is a sign of template incompleteness, and a strong invitation to explore more flexible models, as suggested in Section \ref{Subsec:ann}. 
Indeed, using exponential star formation histories leads to much better results, with a median bias, scatter, and outlier fraction of -0.01, 0.06 and 7.5\% respectively, {\it without applying any zero-point corrections.} On the basis of our results in the previous section, we chose not to apply offsets to these templates. We note that our results are of comparable quality to the ones obtained by the teams participating in the CANDELS photo-z challenge (\citealt{Dahlen2013}) in terms of correspondence between spectroscopic and photometric redshifts, scatter, and outlier fraction in redshift. The sizes of the 68 and 95\% credible intervals are also not far from their nominal value, with 51 and 83\% of objects falling in these two estimated credible intervals. We can apply annealing to obtain more accurate credible regions; sampling at a temperature T = 3 causes these numbers to shift to 65 and 94\% respectively, while leaving bias, scatter and OLF basically unaltered, as shown in Table \ref{Tab:data}. 

\subsection{Photometric redshift estimation with EAZY and SED fitting with SpeedyMC}

As a final experiment in the joint photo-z estimation and SED fitting, we considered combining the use of the publicly available photometric redshift code EAZY (\citealt{EAZY}) with SpeedyMC. EAZY is known to produce accurate photometric redshifts, and we wanted to investigate the following two issues:
\begin{itemize}
\item Whether using the probability distributions obtained by EAZY as priors for the photo-z + SED fitting could further reduce the OLF and scatter obtained by using SpeedyMC alone;
\item Whether combining two different sets of templates (the empirical templates used by EAZY and the SPS templates used by SpeedyMC) would introduce a significant bias in the recovered SED fitting parameters.
\end{itemize}
We used EAZY to compute P(z) distributions for the same catalog described above, adjusting the photometric uncertainties to reproduce the 68 and 95\% credible intervals correctly. These P(z) curves were then used as a prior probability distribution for redshift by SpeedyMC in place of the luminosity function prior; in other words, SpeedyMC would sample each redshift value from these probability distributions. The results were slightly better than the case in which SpeedyMC was used by itself, with bias varying from -0.005 to 0.001, and OLF and scatter both decreasing by about 20\%. Since the size of 68 and 95\% estimated credible regions were still underestimated, we used annealing at T = 3 to correct the size of the error bars. Once again, annealing did not impact negatively the performance of the photo-z + SED fitting, as shown in Table \ref{Tab:data}. 

We were also able to verify that the mean values of SED fitting parameters didn't change significantly whether they were estimated using SpeedyMC (\ie using consistently the same set of templates) or using EAZY to determine the P(z) first, and ``feeding" them into SpeedyMC as a prior. The median difference (percentage of objects for which the difference was larger than the 68\% error) in stellar mass, stellar age, redshift, and E(B-V) in the two cases were found to be -0.01 (12\%), -0.005 (9\%), -0.006 (13\%), 0.003 (4\%). The main driver of the disagreement is the different estimate of redshift, which immediately translates into a different estimate in mass because of the change in luminosity distance. Of course, there is no telling in general which one is correct, but the rule is that if the redshift estimate agrees, the estimates of the other parameters also agree. These results are also in agreement with the similar analysis conducted in \cite{Finkelstein2010}.

We also compared the performance of SpeedyMC with annealing to that obtained by using EAZY \citep{EAZY} together with the publicly available SED fitting code FAST \citep{FAST}. 
We used FAST with input values as close as possible to those used by SpeedyMC (\eg using the same SPS models, initial mass function, dust law, and SFH); however, some differences persist, such as the inclusion of the nebular emission contribution is SpeedyMC, which is not available in FAST.
The performance of these two combined codes is similar to that obtained when SpeedyMC was used by itself, 
and slightly worse than that obtained by using SpeedyMC with EAZY, with median bias of -0.004 (vs 0.001), an outlier fraction of 8.9\% (vs 6.3\%) and similar scatter of 0.05. The difference in the estimate of SED fitting parameters from EAZY + FAST and EAZY + SpeedyMC was also examined, and found to be within the 68\% uncertainties in 90-95\% of cases for all parameters. The size of 68 and 95\% credible regions were estimated correctly by FAST once the outliers were eliminated.

\begin{table}
\begin{center}
\resizebox{\linewidth}{!}{
\begin{tabular}{|l|ccccc|}
\hline

 & \% of objects & \% of objects & & & \\
\hspace{2cm} Settings & with photo-z & with photo-z & average bias & scatter & OLF  \\
& in 68\% region & in 95\% region & in photo-z & in photo-z & in photo-z \\ 
\hline
Data, CSF, SpeedyMC & 42 & 66 & -0.03& 0.06 & 0.16 \\

Data, ESF, SpeedyMC & 51 & 83 & -0.005 & 0.06 & 0.075 \\

Data, ESF, SpeedyMC, T = 3 & 65 & 94 & -0.005 & 0.06 & 0.079 \\

Data, ESF, EAZY + SpeedyMC  & 44 & 74 & 0.002 & 0.049 & 0.057 \\

Data, ESF, EAZY + SpeedyMC, T = 3 & 60 & 93 & 0.001 & 0.049 & 0.063 \\

Data, ESF, EAZY + FAST & 67 & 93 & -0.004 & 0.046 & 0.089 \\

\hline
\end{tabular}}
\caption{Comparison of results for photoz + SED fitting runs on CANDELS data. The choice of SFHs and annealing temperatures are driven by the lessons learned on simulated catalogs. In particular, we confirm on data the following findings: 1. Using more flexible SFHs improves the fitting performance, and 2. Annealing doesn't significantly increase the values of bias, OLF and scatter.}
\label{Tab:data}
\end{center}
\end{table}

\section{Conclusions}

We have investigated the relationship between photometric redshift estimation and SED fitting and assessed performance in both cases through bias, scatter, outlier fraction, and whether the uncertainties were evaluated correctly. We concluded that the performance of photometric redshift determination is strongly correlated with that of SED fitting, and that because the former can be tested on data by analyzing validation samples with spectroscopic redshifts, the latter can be improved accordingly. We tested our hypothesis on various types of mock catalogs, making sure we included severely problematic (but rather common) sources of systematic errors in the photo-z+SED fitting setup. Our main conclusions are:
\begin{itemize}
\item ``Non-algorithmic" (i.e., code-independent) systematics, such as catastrophic failures in the photometry or template incompleteness, will result in an underestimation of the uncertainties both in photo-zs and SED fitting parameters.
\item The presence and severity of these systematics is signaled by a high fraction of redshift outliers. 
\item The performance of the joint SED fitting-photoz estimation can be significantly improved by using more flexible SPS models and more realistic star formation histories.
\item In the presence of systematics, the median stellar age of a galaxy is a more robust parameter than the time elapsed from the beginning of star formation.
\item We could not reach a conclusion about whether adding offsets to the photometry, as customary in photometric redshift codes, might be detrimental to the estimation of SED fitting parameters, but we did not find evidence that it is beneficial.
\item Even when severe non-algorithmic systematics are present, the extent to which uncertainties in redshift and other SED fitting parameters are underestimated is strongly correlated. Therefore, if uncertainties in photo-zs can be recalibrated by using a validation sample with spectroscopic redshifts, those in SED fitting parameters can also be corrected.
\item  Annealing (running MCMC chains at higher temperature, by substituting the likelihood $\cal{L}$ with ${\cal L}^{1/T}$) is a valid method of calibrating the joint uncertainties on photo-zs and other SED fitting parameters without sacrificing accuracy.
\item Using the same code and set of templates for photometric redshift estimation and SED fitting has the advantage of internal consistency and allows one to reconstruct the full multidimensional probability distribution. However, we found that photometric redshift codes such as EAZY can also be used in combination with SED fitting codes. We obtained a competitive performance by using the P(z) from EAZY as a prior in the SED fitting, and using annealing to re-size the uncertainties. The difference in the estimation of SED parameters in the two cases was within the uncertainty in the vast majority of cases.

\end{itemize} 

\section*{Acknowledgments}

It is a pleasure to thank Stefano Andreon, Julia Avez, Guillermo Barro, Gabe Brammer, Steve Finkelstein, Mara Salvato, and the anonymous referee for many useful conversations and suggestions. We acknowledge the Center for Theoretical Physics of the New York City College of Technology for providing computational resources. A.R. acknowledges funding from the P2IO LabEx (ANR-10-LABX-0038) in the framework ``Investissements dÕAvenir" (ANR-11-IDEX-0003-01) managed by the French National Research Agency (ANR), the support of the French ANR under the reference ANR10-BLANC-0506-01-Projet VIRAGE, and financial contribution from the agreement ASI-INAF I/009/10/0 and from Osservatorio Astronomico di Brera. This material is based upon work supported by the National Science Foundation under Grant No. AST-1055919. \\

\end{document}